\documentclass[conference]{IEEEtran}
\IEEEoverridecommandlockouts
\usepackage{cite}
\usepackage{amsmath,amssymb,amsfonts}
\usepackage{algorithmic}
\usepackage{algorithm}
\usepackage{graphicx}
\usepackage{textcomp}
\usepackage{xcolor}
\usepackage{makecell}
\def\BibTeX{{\rm B\kern-.05em{\sc i\kern-.025em b}\kern-.08em
    T\kern-.1667em\lower.7ex\hbox{E}\kern-.125emX}}

\begin{document}

\title{Quantum Approximate Optimization Algorithm for Maximum Likelihood Detection in Massive MIMO\\

\thanks{Corresponding Authors: Zaichen Zhang and Xutao Yu}
}

\author{\IEEEauthorblockN{Yuxiang Liu\IEEEauthorrefmark{1}\IEEEauthorrefmark{4}, 
Fanxu Meng\IEEEauthorrefmark{5}, 
Zetong Li\IEEEauthorrefmark{2}\IEEEauthorrefmark{4},
Xutao Yu\IEEEauthorrefmark{2}\IEEEauthorrefmark{3}\IEEEauthorrefmark{4}, and  
Zaichen Zhang\IEEEauthorrefmark{1}\IEEEauthorrefmark{3}\IEEEauthorrefmark{4}}  

\IEEEauthorblockA{\IEEEauthorrefmark{1}National Mobile Communications Research Laboratory, Southeast University, Nanjing 210096, China}  
\IEEEauthorblockA{\IEEEauthorrefmark{2}State Key Laboratory of Millimeter Waves, Southeast University, Nanjing 210096, China} 
\IEEEauthorblockA{\IEEEauthorrefmark{3}Purple Mountain Laboratories, Nanjing 211111, China}  
\IEEEauthorblockA{\IEEEauthorrefmark{4}Frontiers Science Center for Mobile Information Communication and Security, Southeast University, Nanjing 210096, China} 
\IEEEauthorblockA{\IEEEauthorrefmark{5}College of Artificial Intelligence, Nanjing Tech University, Nanjing 211800, China}  
\IEEEauthorblockA{Corresponding Authors: Zaichen Zhang and Xutao Yu}  
\IEEEauthorblockA{Emails: \{zczhang, yuxutao\}@seu.edu.cn}  
}

\maketitle

\begin{abstract}
In the massive multiple-input and multiple-output (Massive MIMO) systems, the maximum likelihood (ML) detection problem is NP-hard and becoming classically intricate with the number of the transmitting antennas and the symbols increasing. The quantum approximate optimization algorithm (QAOA), a leading candidate algorithm running in the noisy intermediate-scale quantum (NISQ) devices, can show quantum advantage for approximately solving combinatorial optimization problems. In this paper, we propose the QAOA based the maximum likelihood detection solver of binary symbols. In proposed scheme, we first conduct a universal and compact analytical expression for the expectation value of the $1$-level QAOA. Second, a bayesian optimization based parameters initialization is presented, which can speedup the convergence of the QAOA to a lower local minimum and improve the probability of measuring the exact solution. Compared to the state-of-the-art QAOA based ML detection algorithm, our scheme have the more universal and compact expectation value expression of the $1$-level QAOA, and requires few quantum resources and has the higher probability to obtain the exact solution.
\end{abstract}


\section{Introduction}

The massive multiple-input and multiple-output (Massive MIMO) \cite{1}, the core technology of the fifth generation mobile communication system (5G), has emerged as the primary focus of research in both academia and industry. Compared to the classical MIMO technology, by equiping with more antennas in the receiver and transmitter, the massive MIMO can provide some significant advantages, consisting of improving the number of cell service customers and system spectral efficiency, conserving energy and so on. However, the massive MIMO technology also poses challenges to some algorithms in the classical MIMO. Among these algorithms is the well-known ML detection \cite{2,3,4,5},  which is the theoretically optimal detection algorithm but is NP-hard \cite{6} and has the complexity growing exponentially with the number of the antennas and the symbols transmitted. This motives us to explore a new computing paradigm to speedup the ML detection in the massive MIMO while maintaining optimal detection performance.

Quantum computing is a rapidly emerging computing paradigm that offers the potential of speedup over classical computation for certain problems, such as integer factoring \cite{7}, database searching \cite{8} and linear regression \cite{9,10,11,12}, as well as variety of machine-learning tasks \cite{13,14,15,16}. Recently, the researches of quantum computing in wireless communication system have began to attract more attention, consisting of the quantum computing assisted indoor localisation \cite{17}, the optimal routing \cite{18}, multi-user detection \cite{19,20,21} and direction-of-arrival (DOA) estimation \cite{22,23}. However, these algorithms with provable speedups require currently unavailable fault-tolerant quantum computers. Fortunately, we are now entering a pivotal new era, noisy intermediate-scale quantum (NISQ) era, where the availability of the NISQ devices is regarded as a significant step towards the development of a more powerful quantum computer while the number of qubits, gate fidelity, and coherence time are limited. The quantum approximate optimization algorithm (QAOA) \cite{24} which is proposed for approximately solving combinatorial optimization problems, has been widely considered as a promising approach for demonstrating NISQ application. The basic flow of a $p$-level QAOA is to estimate the cost function (the expectation value of the problem Hamiltonian) by sampling from the parameterized quantum circuit composed of $p$ layers of alternating operators with a problem Hamiltonian and a problem-independent mixer Hamiltonian, and call for classical optimizer to iteratively find the optimal parameters for minimizing the cost function. The solution of the problem as a bitstring can be obtained by sampling the output of the parameterized quantum circuit with the optimal parameters.

The milestone work on connecting the ML detection with the QAOA is proposed in the Ref \cite{25}, where authors transform the model of the ML detection into a problem Hamiltonian encoding the solution as the ground state, derive the analytical expression of the cost function of the $1$-level QAOA, and optimize the cost function using the COBYLA optimizer with the random parameters initialization. However, in this work, the analytical expectation expression of the $1$-level QAOA is not universal and compact enough especially when the problem size $N$ is greater than $3$. Moreover, as the cost function of the QAOA is nonconvex \cite{26}, the optimization with the random parameters initialization usually converges to some low-quality nondegenerate local optimas \cite{27}.

Therefore, in this paper, we propose an improved QAOA based ML detection in the massive MIMO. Our main contributions are summarized as follows. First, we derivate a more compact and universal analytical expression of the cost function of the $1$-level QAOA. Second, a  bayesian optimization based parameters initialization scheme is presented, which can learn the high-quality initial parameters by optimizing a set of small-scale and classically simulable problem instances using the bayesian optimization algorithm. Our initialization scheme can speedup the convergence of the cost function to a lower local minimum as well as improve the probability of measuring the exact solution. Last, we provide a series of numerical results to demonstrate the advantage of our algorithm over the state-of-the-art scheme.

The paper is organized as follows. In Section \ref{BACKGROUND}, we introduce the background about quantum gates, review the QAOA, the ML detection and bayesian optimization algorithm. In Section \ref{QAOA}, we detail the universal analytical expectation of the $1$-level QAOA, the bayesian optimization based parameters initialization. In Section \ref{NE}, the numerical results are presented to demonstrate the advantage of the our algorithm. Finally, a discussion and summary are presented in Section \ref{Conclusion}.

\section{BACKGROUND}
\label{BACKGROUND}
\subsection{Quantum Computing}
Quantum gate, as a unitary matrix, transforms a quantum state to another and preserves the norm of the quantum state. The basic quantum gates can be split into two groups, single-qubit gates and two-qubits gates. The commonly used single-qubit gates are Pauli gates ($I$, $\sigma _{x}$, $\sigma _{y}$ and $\sigma _{z}$) and Hadamard gate ($H$), which can respectively be denoted as the following unitary matrices,
\begin{align}
  I=\begin{pmatrix}
  1&0 \\
  0&1
\end{pmatrix}
&&\sigma _{x} =\begin{pmatrix}
  0& 1\\
  1&0
\end{pmatrix}
&& \sigma _{y} =\begin{pmatrix}
  0& -i\\
  i&0
\end{pmatrix}
\end{align}
\begin{align}
  \sigma _{z}=\begin{pmatrix}
  1&0 \\
  0&-1
\end{pmatrix}
&&H =\frac{1}{\sqrt{2} } \begin{pmatrix}
  1& 1\\
  1&-1
\end{pmatrix}
\end{align}
Moreover, the Pauli rotation gates are defined as
\begin{equation}
      R_{x} \left ( \theta  \right ) =e^{-i\frac{\theta}{2}\sigma _{x}   }=\begin{pmatrix}
 \cos\frac{\theta}{2}  & -i\sin\frac{\theta}{2} \\
  -i\sin\frac{\theta}{2} & \cos\frac{\theta}{2}
\end{pmatrix}
\end{equation}
 \begin{equation}
    R_{y} \left ( \theta  \right ) =e^{-i\frac{\theta}{2}\sigma _{y}   }=\begin{pmatrix}
 \cos\frac{\theta}{2}  & -\sin\frac{\theta}{2} \\
  \sin\frac{\theta}{2} & \cos\frac{\theta}{2}
\end{pmatrix}
 \end{equation}
\begin{equation}
      R_{z} \left ( \theta  \right ) =e^{-i\frac{\theta}{2}\sigma _{z}   }=\begin{pmatrix}
 e^{-i\frac{\theta}{2}}  & 0 \\
  0 & e^{i\frac{\theta}{2}}
\end{pmatrix}
\end{equation}
The notable two-qubits gates are CNOT gate and SWAP gates as follows
\begin{align}
  CNOT=\begin{pmatrix}
 1 & 0 & 0 & 0\\
 0 & 1 & 0 & 0\\
 0 & 0 & 0 & 1\\
 0 & 0 & 1 &0
\end{pmatrix}
&&
SWAP=\begin{pmatrix}
 1 & 0 & 0 & 0\\
 0 & 0 & 1 & 0\\
 0 & 1 & 0 & 0\\
 0 & 0 & 0 &1
\end{pmatrix}
\end{align}
\subsection{Quantum Approximate Optimization Algorithm}
\begin{figure}[!t]
\centering
\includegraphics[width=3.5in]{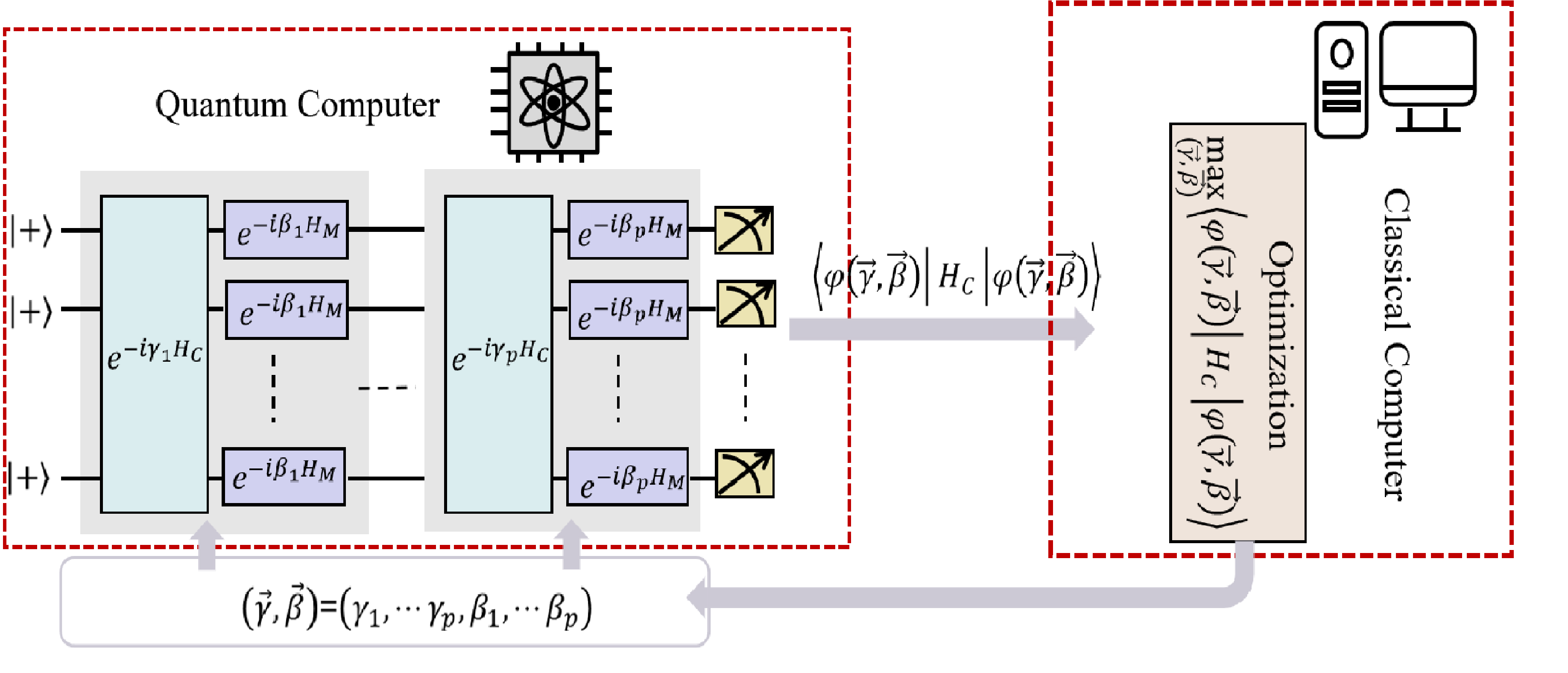}
\caption{Schematic of a $p$-level quantum approximation optimization algorithm.}
\label{fig_QAOA}
\end{figure}

For a combinatorial optimization problem defined on $N$-bit binary strings $\mathbf{z}= z_{1}z_{2}\dots z_{N}$, the goal is to determine a string that maximizes or minimizes a given classical objective function $C\left ( \mathbf{z}  \right ) :\left \{ -1,+1 \right \} ^{N} \mapsto \mathbb{R}$. To solve this problem by the QAOA, the classical objective function can be converted to a quantum problem Hamiltonian by promoting each binary variable $z_{i}$ to a quantum spin $\sigma_{z}^{i}$:
\begin{equation}
    H_{C} =C\left ( \sigma _{z} ^{1},\sigma _{z} ^{2},\dots , \sigma _{z} ^{N}\right )
\end{equation}
For the $p$-level QAOA, which is visualized in Fig.\ref{fig_QAOA}
, the quantum processor is initialized in the state $\left | +  \right \rangle ^{\otimes N} $ and then apply $p$ times the problem Hamiltonian $H_{C}$ and the mixing Hamiltonian $H_{M} = {\textstyle \sum_{j=1}^{N}} \sigma _{x}^{j}$ alternately to generate a variational wave function as follows
\begin{equation}
\left | \psi\left ( \mathbf{\gamma} ,\mathbf{\beta}   \right )   \right \rangle=e^{-i\beta _{p}H_{M} }e^{-i\gamma _{p}H_{C} }\cdots e^{-i\beta _{1}H_{M} }e^{-i\gamma _{1}H_{C} } \left | +  \right \rangle ^{\otimes N}
\end{equation}
which is parameterized by $2p$ variational parameters $\gamma_i$ and $\beta_i$ $(i=1,2,\cdots, p)$. Then, the expectation value of $H_C$ in this variational state is calculated as
\begin{equation}
C_{p}=\left \langle \psi \left ( \mathbf{\gamma},\mathbf{\beta}     \right ) \left |  H_C\right |  \psi \left (  \mathbf{\gamma},\mathbf{\beta} \right )  \right \rangle
\end{equation}
which is determined by repetitive measurements on the quantum system in the computational basis. The classical computer is then applied to explore the optimal parameters  $\mathbf{\gamma}  ^{\ast }$ and $\mathbf{\beta}  ^{\ast }$ so as to maximize the expectation value $C_{p}$:
\begin{equation}
\mathbf{\gamma}  ^{\ast } ,\mathbf{\beta}  ^{\ast }=\max_{\mathbf{\gamma},\mathbf{\beta}} C_p\left ( \mathbf{\gamma},\mathbf{\beta} \right )
\end{equation}
The solution of the combinatorial optimization problem can be determined by measuring the quantum state $\left | \psi \left ( \mathbf{\gamma^{\ast } },\mathbf{\beta^{\ast } }   \right )  \right \rangle $ in the computational basis.

\subsection{Maximum Likelihood Detection}
The ML detection can theoretically provide the best performance in all the detection algorithms. Given a channel matrix $H\in \mathbb{R}^{N_{r}\times N_{t}  }$ and a received signal $\mathbf{y } \in \mathbb{R} ^{N_{r} }$ where ${N_{r}}$ and ${N_{t}}$ are the number of the receiving and transmitting signals, the core of the ML detection is to explore all the combination of transmitting signals to search an optimal signal $\mathbf{ \tilde{x} }$ satisfying the following objective:
\begin{equation}
\mathbf{ \tilde{x} } =\min_{\mathbf{x} \in A^{N_{t} } } \left \| \mathbf{y} -H\mathbf{x}  \right \| _{2}
\end{equation}
where $A^{N_{t} } =\left \{ -1,1 \right \} ^{N_{t}}$ is the set of all possible binary symbols transmitted. Therefore, the complexity of the ML detection with binary symbols grows exponentially with the number of the transmitting antennas.

\subsection{Bayesian optimization algorithm}
The bayesian optimisation algorithm utilizes a surrogate model of the global cost function to guide the optimization.The surrogate model at unevaluated $\gamma$ or $\beta$ points is inferred based on all previous cost functions evaluations by the means of the bayes rule. Every optimization round, the next $\gamma$ or $\beta$ point
is determined by maximizing a acquisition function over the surrogate model. The surrogate model is usually the gaussian process. Acquisition functions are heuristics employed to evaluate the usefulness of one of more design points for achieving the objective of maximizing the underlying black box function. In this paper, we use the following the upper confidence bound (UCB) as the the acquisition function:
\begin{equation}
UCB\left ( \gamma _t,\beta _t \right ) = \mu \left ( \gamma, \beta \right ) +\kappa \sigma \left ( \gamma, \beta \right )
\end{equation}
where $\mu \left ( \gamma, \beta \right )$ and $\sigma^2 \left ( \gamma, \beta \right )$ are the mean and variance functions of the Gaussian process surrogate model. The hyperparameter $\kappa$ is chosen to tradeoff the exploitation and exploration.
The detailed flow of the bayesian optimization algorithm is presented in the Algorithm 1:
\begin{algorithm}[H]
\caption{Bayesian optimization algorithm.}\label{alg:alg1}
\begin{algorithmic}
\STATE {for t = 1,2,... do}
\STATE \hspace{0.2cm} 1.Find ($\gamma_t, \beta_t$) by optimizing the acquisition function:$\left ( \gamma_t, \beta_t \right ) =\mathrm{arg} \max_{\left ( \gamma_t, \beta_t \right )}U\left (\gamma_t, \beta_t|D_{1:t-1}  \right ) $;
\STATE \hspace{0.6cm}2. Sample the objective function:$y_{t}=-C_{p}\left ( \gamma _t, \beta _t \right ) + \varepsilon _t$;
\STATE \hspace{0.6cm}3.Augment the data $D_{1:t}=\left \{ D_{1:t-1},\left \{ \left ( \gamma_t,\beta _t  \right ) ,y_t \right \}  \right \} $.
\STATE end for
\end{algorithmic}
\label{alg1}
\end{algorithm}

\section{QUANTUM APPROXIMATE OPTIMIZATION ALGORITHM FOR Maximum Likelihood Detection DETECTION}
\label{QAOA}
In this section, we first introduce the transformation between the model of the ML detection and the problem Hamiltonian. Then, we present the universal and compact analytical expression of the expectation value of the problem Hamiltonian. Finally, the detailed bayesian optimization based parameters initialization subroutine is depicted.
\subsection{Model Transformation}
Here, similarly to the ref \cite{25}, we conduct the transformation between the model of the ML detection and the problem Hamiltonian. The detailed procedure is given as follows:

1. Rewrite the model of the ML detection as
\begin{equation}
\begin{array}{c}
  \min_{x\in \left \{ -1,+1 \right \} ^{N_t} } \left \| y-Hx \right \| _{2} \\
  \Longleftrightarrow \min_{x\in \left \{ -1,+1 \right \} ^{N_t} }\left \| y-Hx \right \| _{2}^{2}\\
\Longleftrightarrow \min_{x\in \left \{ -1,+1 \right \} ^{N_t} }y^Ty-2 y^THx  +x^TH^THx
\\ \Longleftrightarrow\min_{x\in \left \{ -1,+1 \right \} ^{N_t} } {\textstyle \sum_{i<j}^{N_t}} 2A_{ij}x_ix_j-{\textstyle \sum_{k=1}^{N_t}} 2b_kx_k
\end{array}
\end{equation}
where $A=H^TH$ is a symmetrical matrix and $b=y^TH$.

2. Promote each binary variable $x_{i}$ to a quantum spin $\sigma_{z}^{i}$ to obtain the following problem Hamiltonian
\begin{equation}
H_C={\textstyle \sum_{i<j}^{N_t}} 2A_{ij}\sigma ^i_z\sigma ^j_z-{\textstyle \sum_{k=1}^{N_t}} 2b_k\sigma ^k_z
\end{equation}
whose ground state is corresponding to the solution of the original ML detection.

\subsection{Expectation Value of the $1$-level QAOA}
\begin{figure}[!t]
\centering
\includegraphics[width=3.5in]{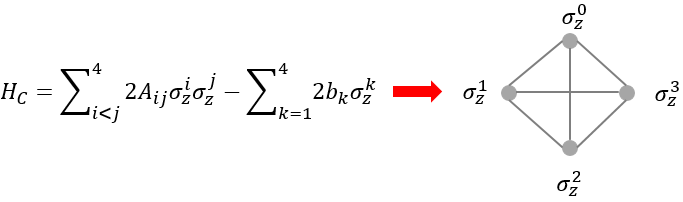}
\caption{An undirected graph corresponding to the problem Hamiltonian with $N_{t}=4$.}
\label{fig_graph}
\end{figure}

In this subsection, we present the universal and compact expectation expression of the $1$-level QAOA.  For the expectation value of the problem Hamiltonian
\begin{equation}
\begin{array}{c}
C_p\left ( \gamma ,\beta \right ) =\left \langle \psi \left ( \gamma ,\beta  \right )\left | H_C \right | \psi \left ( \gamma ,\beta  \right ) \right \rangle =
\\
{\textstyle \sum_{i<j}^{N_t}} 2A_{ij}\left\langle \psi \left ( \gamma ,\beta  \right )  \left |\sigma ^i_z\sigma ^j_z  \right | \psi \left ( \gamma ,\beta  \right ) \right \rangle \\
-{\textstyle \sum_{k=1}^{N_t}} 2b_k\left \langle \psi \left ( \gamma ,\beta  \right )\left | \sigma ^k_z \right | \psi \left ( \gamma ,\beta  \right ) \right \rangle
\end{array}
\end{equation}
, we first characterize the problem Hamiltonian as an undirected graph $G=(V,E)$, where any vertices $i$ represents a $\sigma ^i_z$ and any edge $e_{ij}$ represents a term $\sigma ^i_z\sigma ^j_z$ as visualized in Fig.\ref{fig_graph}. Thus, the expectation value can be rewritten as
\begin{equation}
\begin{array}{l}
C_p\left ( \gamma ,\beta \right ) = \sum_{\left ( i,j \right ) \in E}^{} 2A_{ij}\left\langle \psi \left ( \gamma ,\beta  \right )  \left |\sigma ^i_z\sigma ^j_z  \right | \psi \left ( \gamma ,\beta  \right ) \right \rangle \\
-\sum_{k \in V}^{} 2b_k\left \langle \psi \left ( \gamma ,\beta  \right )\left | \sigma ^k_z \right | \psi \left ( \gamma ,\beta  \right ) \right \rangle
\end{array}
\end{equation}
Then, we apply the causal cone to conduct the terms $C_{1}^{ij}=\left\langle \psi \left ( \gamma ,\beta  \right )  \left |\sigma ^i_z\sigma ^j_z  \right | \psi \left ( \gamma ,\beta  \right ) \right \rangle$ and $C_{1}^{i}=\left \langle \psi \left ( \gamma ,\beta  \right )\left | \sigma ^i_z \right | \psi \left ( \gamma ,\beta  \right ) \right \rangle $ as follows
\begin{equation}
\begin{array}{l}
 C^i_{1}=- \sin (2 \beta) \sin \left(4 \gamma b_{i}\right) \prod_{\left ( i,k \right ) \in E}  \cos \left(4 \gamma A_{i k}\right)
\end{array}
\end{equation}
\begin{equation}
\begin{array}{l}
C^{i j}_1= \\
\frac{1}{2} \sin (4 \beta) \sin \left(4A_{i j} \gamma \right)\cos \left(4b_{j} \gamma \right)\prod_{\substack{  k\ne i
}} \cos \left(4A_{j k} \gamma \right)+\\
\frac{1}{2} \sin (4 \beta) \sin \left(4A_{i j \gamma }\right)\cos \left(4 b_{i}\gamma \right)
 \prod_{\substack{k\ne j
}} \cos \left(4 \gamma A_{i k}\right) -\\
\frac{1}{2}(\sin (2 \beta))^{2}(cos(4A_{ij}\gamma))^2 \cos \left(4 \gamma\left(b_{i}+b_{j}\right)\right)
\times \\ \prod_{k\ne i,k\ne j} \cos \left(4 \gamma\left(A_{i k}+A_{j k}\right)\right)+\\
\frac{1}{2}(\sin (2 \beta))^{2}(cos(4A_{ij}\gamma))^2\cos \left(4 \gamma\left(b_{j}-b_{i}\right)\right)\times \\ \prod_{k\ne i,k\ne j} \cos \left(4 \gamma\left(A_{j k}-A_{i k}\right)\right)
\end{array}
\end{equation}
Compared to the result in the state-of-the-art \cite{25} as
\begin{equation}
\begin{array}{l}
 C^i_{1}=
\left\langle+^{3}\right| U^{\dagger}\left(A_{k^{\prime}, k}, \gamma\right) U^{\dagger}\left(A_{k, k^{\prime \prime}}, \gamma\right) \sigma_{z}^{(k)} U^{\dagger}\left(b_{k}, \beta\right) \sigma_{x}^{k} \\
U\left(b_{k}, \beta_{1}\right) U\left(A_{k, k^{\prime \prime}}, \gamma\right) U^{\dagger}\left(A_{k^{\prime}, k}, \gamma\right) \sigma_{z}^{(k)}\left|+{ }^{3}\right\rangle
\end{array}
\end{equation}
\begin{equation}
\begin{array}{l}
C^{i j}_1=  \left\langle+^{N}\right| U_{N}^{\dagger}(\gamma, \beta) \cdots U_{1}^{\dagger}(\gamma, \beta) \sigma_{z}^{(k)} \sigma_{z}^{(l)} U_{1}(\gamma, \beta) \cdots \\
\quad U_{N}(\gamma, \beta)\left|+{ }^{N}\right\rangle
\end{array}
\end{equation}
where $U_{k}(\gamma, \beta)=\prod_{l>k} U\left(A_{k, l}, \gamma\right) U\left(b_{k}, \gamma\right) U(k, \beta)$, and $U\left(A_{k, l}, \gamma\right)=e^{-A_{k, l}\gamma_{z}^{k} \gamma_{z}^{l}} $, $U\left(b_{k}, \gamma\right)=e^{-2b_{k} \gamma\sigma _{z} }$, $ U(k, \beta)=e^{-\beta\sigma _{x}^{k}  } $, and $k^{\prime}<k<k^{\prime \prime}$, obviously, our analytical expression is more compact, universal and concrete.


\subsection{Bayesian optimization based parameters initialization}
\begin{figure}[!t]
\centering
\includegraphics[width=3.5in]{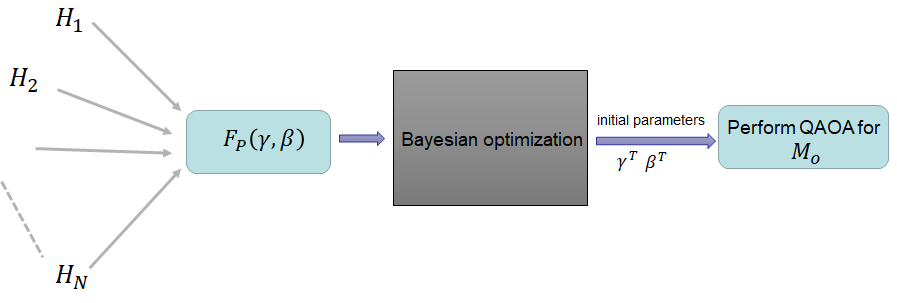}
\caption{Schematic of the bayesian optimization based parameters initialization.}
\label{fig_BO}
\end{figure}

In the QAOA, the complexity of the optimization landscape grows with increasing $p$, meantime, the landscape is also characterized by a large number of local sub-optimal minima. The convergence of classical optimization algorithms into such sub-optimal solutions is demonstrated to be a potential bottleneck. To mitigate the above issue and reduce the consumption of quantum resources, motived by the Meta-VQE in the ref \cite{28}, we present a bayesian optimization based parameters initialization subroutine. Our scheme is data-driven and takes a set of small-scale and classically simulated $p$-level QAOA based ML detection instances as the input to construct the objective function of the bayesian optimization algorithm. By a few optimization rounds, we can learn $2p$ parameters ${(\gamma_i, \beta_i)}^p_{i=1}$, which can be used as the initial parameters for others even larger-scale $p$-level QAOA based ML detection instances. The details of the proposed scheme is shown in Algorithm 2 and Fig.\ref{fig_BO}.
\begin{algorithm}[H]
\caption{Bayesian optimization based parameters initialization.}\label{alg:alg2}
\begin{algorithmic}
\STATE \textbf{Input:} The set of small-scale ML detection problem instances $\left \{ M_i \right \} _{i=1}^{N}$ 
\STATE 1. Construct problem Hamiltonians set $\left \{ H_i \right \} _{i=1}^{N}$ for $\left \{ M_i \right \} _{i=1}^{N}$.
\STATE 2. Calculate every expectation value $C_{p}^{H_i} \left ( \gamma ,\beta \right )  =\left \langle \psi \left ( \gamma ,\beta \right ) \left |  H_i\right |\psi \left ( \gamma ,\beta \right )   \right \rangle $ for every $H_i$ with $2p$ parameters.
\STATE 3. Construct the objective function $F_P\left ( \gamma ,\beta  \right )=\frac{1}{N} {\textstyle \sum_{i=1}^{N}}C_{p}^{H_i}  \left ( \gamma ,\beta  \right )  $.
\STATE 4. \textbf{If } Train, \textbf{then}:
\STATE \hspace{0.4cm} Perform bayesian optimization $T$ times with the
\STATE \hspace{0.4cm} objective function $F_P\left ( \gamma ,\beta  \right )$ to obtain initial parameters
\STATE \hspace{0.4cm}  $\gamma^T$ and $\beta^T$.
\STATE \hspace{0.2cm} \textbf{End If}
\STATE 5. \textbf{End}
\end{algorithmic}
\label{alg2}
\end{algorithm}
Then, for any other even a large-scale ML detection instance $M_{o}$, we can perform the classical optimization with the initial parameters $\gamma^T$ and $\beta^T$.

\section{NUMERICAL EXPERIMENT}
\label{NE}
In this section, we conduct some numerical experiments to show the advantage of the proposed scheme over the state-of-the-art results.

\begin{figure}[!t]
\centering
\includegraphics[width=3.2in]{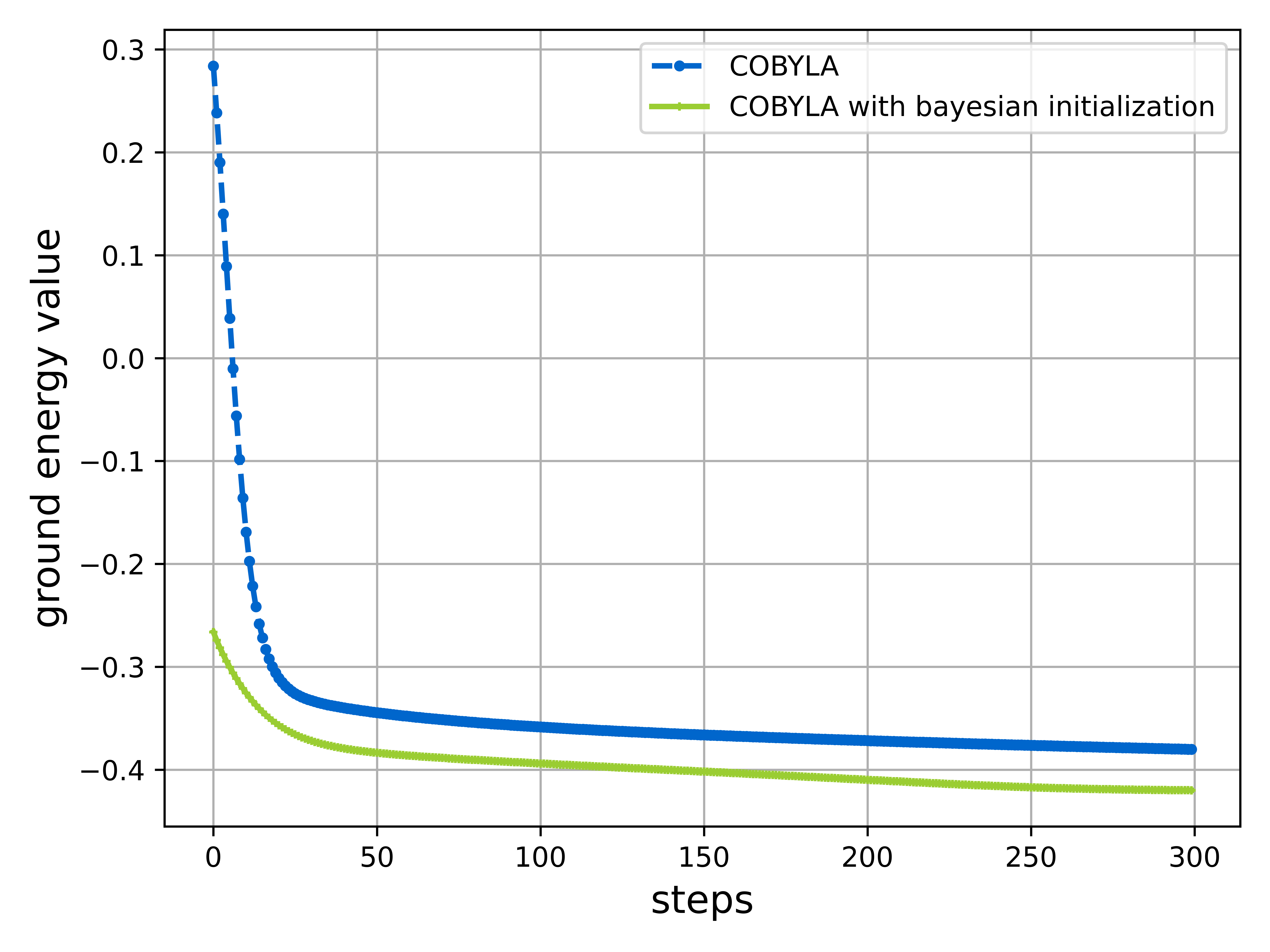}
\caption{The convergence curve of the cost function using COBYLA optimizer with and without parameters initialization.}
\label{fig_9}
\end{figure}

\begin{figure}[!t]
\centering
\includegraphics[width=3.5in]{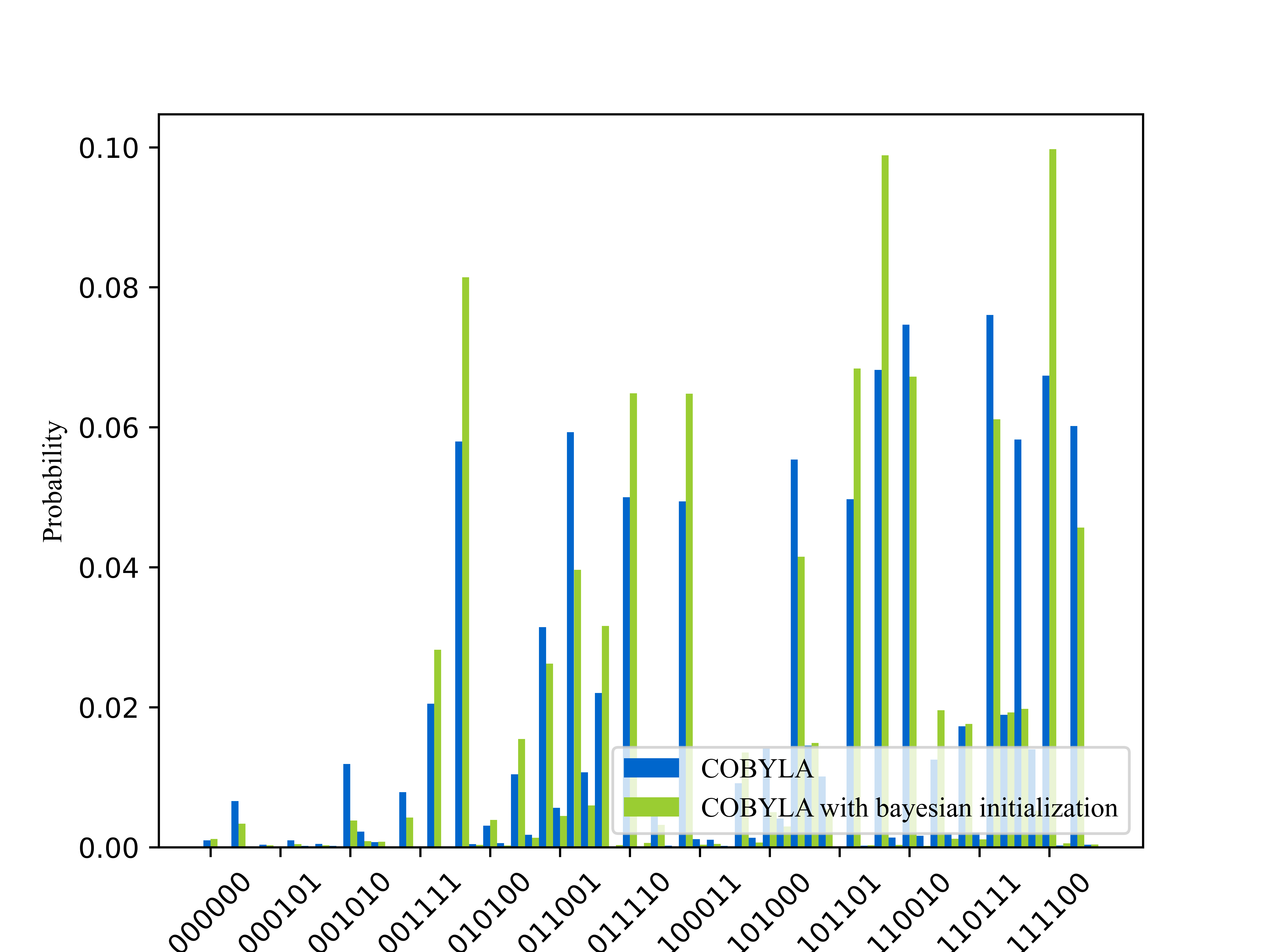}
\caption{The probability measuring the states using COBYLA optimizer with and without parameters initialization.}
\label{fig_10}
\end{figure}

To demonstrate the advantage of our bayesian optimization based parameters initialization, we first randomly select $100$ ML problem instances with $N_{t} \in \left \{ 2,3\right \}$, where the MIMO channel $H$ and the noise $n$ are chosen as independent and identically distributed, zero-mean, real-valued
normal random variables, i.e. $H\sim \mathcal{N} \left ( 0,1 \right )$ and $n\sim \mathcal{N} \left ( 0,1 \right )$. Second, the set of the problem Hamiltonian corresponding to the above problem instances is constructed based on the Equation (13). Next, we construct the objective function with $2p=6$ paramters based on the Step 3 in Algorithm \ref{alg2} and perform $T=10$ rounds of the bayesian optimization using the bayes-opt optimization package. 
Finnally, randomly generating a ML problem instances with $N_t=6$, where $H\sim \mathcal{N} \left ( 0,1 \right )$, $n\sim \mathcal{N} \left ( 0,1 \right )$ and the transmitting signal is $x=\left [ -1, -1, -1, -1,  1,  1 \right ] ^{T}$, we optimize the corresponding cost function by the COBYLA optimizer using the parameters obtained after the bayesian optimization as the initial parameters. As shown in Fig.\ref{fig_9}, our scheme can converge to a lower local minimum. As shown in Fig.\ref{fig_10}, the proposed scheme measures the state $\left | 111100  \right \rangle $ with the highest probability $0.0997325599531442$, where the state $\left | 111100  \right \rangle $ corresponds exactly to the transmitting signal $x=\left [ -1, -1, -1, -1,  1,  1 \right ] ^{T}$ by the following mapping
\begin{equation}
\begin{array}{l} 
 \sigma _{z}^{1} \left | 1  \right \rangle =(-1)\left | 1  \right \rangle\longrightarrow -1; \sigma _{z}^{2} \left | 1  \right \rangle =(-1)\left | 1  \right \rangle\longrightarrow -1;\\
\sigma _{z}^{3} \left | 1  \right \rangle =(-1)\left | 1  \right \rangle\longrightarrow -1; \sigma _{z}^{4} \left | 1  \right \rangle =(-1)\left | 1  \right \rangle\longrightarrow -1;\\
\sigma _{z}^{5} \left | 0  \right \rangle =(1)\left | 0  \right \rangle\longrightarrow 1; \sigma _{z}^{6} \left | 0  \right \rangle =(1)\left | 0  \right \rangle\longrightarrow 1;\\
\end{array}
\end{equation}
however, COBYLA optimizer without parameters initialization measures the state $\left | 111100  \right \rangle $ with the probability $0.06738554612413077$     and the state $\left | 111000  \right \rangle $ with the highest probability $0.07605201170977091$.

\section{Conclusion}
\label{Conclusion}
In the present study, we proposed an improved QAOA based ML detection. In our scheme, we first derive the universal and compact analytical expression of the expectation value of the $1$-level QAOA. Second, the bayesian optimization based parameters initialization is proposed to improve the convergence of the cost function optimization and the probability measuring the exact solution. By the theoretical proof and some numberical experiments, we can demonstrate the advantage of the proposed scheme over the the state-of-the-art schemes.

\section*{Acknowledgments}
This work is supported by NSFC projects (61871111 and 61960206005) and the Fundamental Research Funds for the Central Universities 2242022k60001.

\section{Simple References}

\end{document}